\documentclass[aps, prd, 12pt, eqsecnum, tightenlines, notitlepage, nofootinbib, floatfix]{revtex4-1}
\pdfoutput=1

\PassOptionsToPackage{unicode}{hyperref}

\usepackage{hyperref, graphicx, slashed, xcolor, amsmath, feynmp,float}
\allowdisplaybreaks

\begin{document}

\normalsize
\vspace{-0.5cm}
\begin{flushright}
CALT-TH-2019-010\\
\vspace{1cm}
\end{flushright}

\title{New Shapes of Primordial Non-Gaussianity from Quasi-Single Field Inflation with Multiple Isocurvatons}

\author{Michael McAneny and Alexander K. Ridgway}
\affiliation{Walter Burke Institute for Theoretical Physics, California Institute of Technology, Pasadena, CA 91125}

\begin{abstract}
We study a simple extension of quasi-single field inflation in which the inflaton interacts with multiple extra massive scalars known as isocurvatons.  Due to the breaking of time translational invariance by the inflaton background, the theory includes kinetic mixings among the inflaton and isocurvatons.  These mixings give rise to novel new features in the primordial non-Gaussianities of the scalar curvature perturbation.  A noteworthy feature is the amplitude of the squeezed bispectrum can grow nearly as $(k_s/k_l)^{-3}$ while oscillating as ${\rm \cos\gamma \log}(k_s/k_l)$, where $k_s/k_l$ is the ratio of the lengths of the short and long wavevectors.  Observation of such a shape would provide evidence for the existence of multiple isocurvatons during inflation.  In addition, we consider the effects of these non-Gaussianities on large-scale structure.     
\end{abstract}

\maketitle

\section{Introduction}

The inflationary paradigm \cite{SKS} posits a period of time shortly after the big bang during which the universe's energy density was dominated by vacuum energy and the size of the universe grew exponentially.  Such an era would explain the near isotropy of the CMB and the near flatness of the universe.  At the same time, inflation provides a quantum mechanical origin for the energy density perturbations which have an almost scale-invariant Harrison-Zel'dovich power spectrum \cite{mshsgpbst} (see \cite{Baumann:2009ds} for a review).

The simplest inflationary theories are known as single-field inflation models and involve a single scalar field called the ``inflaton."  In slow-roll models, the inflaton vacuum expectation value $\phi_0$ is initially displaced from the minima of the inflaton potential $V_{\rm sr}(\phi)$.  The potential is chosen such that there is a period of time (the inflationary era) during which $\phi_{0}$'s potential energy dominates over its kinetic energy and drives the exponential expansion of the universe.  After more than $50-60$ e-folds of expansion, $\phi_{0}$'s kinetic energy becomes large and inflation ends.

Since $\dot{\phi}_0 \neq 0$, time-translational invariance is spontaneously broken.  This gives rise to a Goldstone boson $\pi$ that sources scalar curvature perturbations \cite{Cheung:2007st}.  In the simplest single-field models, the curvature perturbations are approximately Gaussian \cite{Maldacena:2002vr}.  To produce large primordial non-Gaussianities (PNG), one can add extra fields to the inflationary field content and include interactions between $\pi$ and the new fields\footnote{Large non-Gaussianities can also be achieved in more complicated single field models, see for example \cite{Alishahiha:2004eh,Chen:2006xjb}.}.  This leads to interesting new shapes and features of  primordial non-Gaussianity that could be observed in the CMB and large-scale structure (LSS) and used to constrain the inflationary theory.

Quasi-single field inflation (QSFI) is a well-studied extension of single-field inflation models \cite{Chen:2009zp} that could potentially give rise to significant PNG.  In QSFI, one adds a single extra scalar field $\sigma$ of mass $m$, known as the isocurvaton, and includes a kinetic mixing between $\sigma$ and $\pi$ of the form\footnote{Kinetic mixing terms such as $\mu \dot{\pi} \sigma$ can appear in the Lagrangian because the inflaton vev spontaneously breaks Lorentz invariance} $\mu \dot{\pi} \sigma$ and a potential for the isocurvaton $V(\sigma)$.  In QSFI the isocurvaton never fulfills the roll of the inflaton, rather, its purpose is to generate PNG through its interactions with $\pi$.  

The connected three- (bispectrum) and four- (trispectrum) point functions of $\pi$ in QSFI have been studied extensively \cite{Chen:2009zp, Chen:2010xka, Baumann:2011su, Baumann:2011nk, Assassi:2012zq, Noumi:2012vr, An:2017hlx, An:2017rwo}.  The squeezed limit of the bispectrum, which occurs when the lengths of two of the three wavevectors are roughly equal and much larger than the length of the third wavevector, has been shown to exhibit particularly interesting behavior.  Let $k_l$ denote the length of the larger wavevectors and $k_s$ the length of the shorter one.  It has been shown that if $m$ and $\mu$ are both much smaller than the Hubble constant during inflation $H$, then the magnitude of the squeezed bispectrum in QSFI grows approximately as $(k_s/k_l)^{-3}$.  In the opposite limit, $m,\mu \gg H$, the magnitude of the squeezed bispectrum was found to oscillate logarithmically in $k_s/k_l$ but grow only as $(k_s/k_l)^{-3/2}$ \cite{Chen:2009zp, An:2017hlx}.

Since single-isocurvaton QSFI gives rise to distinct PNG, it is worthwhile to study theories that include multiple isocurvatons $\sigma_I$ (we will call these theories multi-isocurvaton QSFI).  It turns out that interactions such as $\rho (\dot{\sigma}_{1} \sigma_{2} - \dot{\sigma}_2 \sigma_1)$ and $\mu_{1} \dot{\pi} \sigma_1$ give rise to novel features in the PNG.  Specifically, for certain choices of $\rho$ and $\mu_{1}$, the squeezed bispectrum undergoes logarithmic oscillations in $k_s/k_l$ and grows approximately as $(k_s/k_l)^{-3}$.  This is very different from the behavior of the squeezed bispectrum in single-isocurvaton QSFI, which can only exhibit oscillations or nearly cubic power law growth in different limits of $\mu$ and $m$.  

Several previously studied inflationary models predict log-oscillating or oscillating shapes for the bispectrum \cite{Chen:2006xjb,Chen:2006nt,Chen:2010xka,Chen:2010bka,Agullo:2010ws,Agullo:2011xv,Adshead:2011jq,Achucarro:2012fd,Achucarro:2014msa,Behbahani:2011it,Arkani-Hamed:2015bza,Flauger:2014ana,Flauger:2016idt,Wang:2018tbf,Bolis:2019fmq}. Using shape templates, the Planck collaboration has begun constraining these oscillating/log-oscillating bispectra \cite{Ade:2015ava}. However, none of these shapes simultaneously exhibit log-oscillations and nearly cubic power law growth in the squeezed limit.  In this paper, we provide a bispectrum shape template for multi-isocurvaton QSFI that could be searched for experimentally.  

Previous work has demonstrated that non-Gaussianities that grow like $(k_s/k_l)^{-3}$ in the squeezed limit can have significant effects on large-scale structure \cite{Allen:1987vq, Dalal:2007cu, Baumann:2012bc, Gleyzes:2016tdh}.  For example, in single-isocurvaton QSFI models with $m,\mu \ll H$, non-Gaussian contributions to the dark matter halo-halo power spectrum $P_{hh}(k_s)$ become much larger than the Gaussian contribution as $k_s\rightarrow 0$.  The halo-halo power spectrum $P_{hh}(q)$ will also be sensitive to the extra fields and interactions present in multi-isocurvaton QSFI.  We will show that for certain choices of the parameters $\rho$ and $\mu_{1}$, the bispectrum and trispectrum contributions to $P_{hh}(k_s)$ will dominate over the Gaussian contribution while oscillating logarithmically as $k_s\rightarrow 0$.    

The purpose of this work is to determine the shapes of the bispectrum and trispectrum in multi-isocurvaton QSFI and explore their effects on LSS.  In section \ref{model and mode functions} we write down a general quadratic Lagrangian for multi-isocurvaton QSFI and determine the mode functions of the $\pi$ and $\sigma_I$ fields.  In section \ref{nongaussianity section} we compute the primordial bispectrum and trispectrum due to a cubic interaction of the form $\sigma_1^3$, and provide a template for the bispectrum shape.  Finally, in section \ref{lls section} we calculate the contribution of the PNG to the halo-halo power spectrum $P_{hh}(k_s)$ and halo-matter power spectrum $P_{hm}(k_s)$.

\section{The Model and Mode Functions}
\label{model and mode functions}
We consider multi-isocurvaton QSFI, which include an inflaton field and $N$ extra scalar fields known as ``isocurvatons."  The inflaton develops a time-dependent vacuum expectation value $\phi_{0}$ that sources a background de-Sitter metric
\begin{align}\label{de Sitter}
    ds^2 = \frac{1}{(H\tau)^2}\left
    (d\tau^2 - (dx^i)^2\right)
\end{align}
where $H$ is the Hubble parameter during inflation and $\tau$ is proper time.  We assume that $\phi_{0}$ exhibits a slow-roll trajectory, which means $\ddot{\phi}_0 \simeq 0$ and $H$ is approximately constant throughout the inflationary era (see \cite{Baumann:2009ds} for a review of slow-roll inflation).  The quantum fluctuations of the inflaton and isocurvaton fields perturb the metric about (\ref{de Sitter}) and source scalar and tensor curvature fluctuations, $\zeta(\tau,\vec{x})$ and $\gamma_{ij}(\tau,\vec{x})$.

To describe the field fluctuations, we use the effective field theory of inflation formalism \cite{Cheung:2007st}.  Following \cite{Cheung:2007st}, we choose uniform inflaton gauge in which the inflaton fluctuations are set to zero.  We then observe that the time-dependent inflaton vev spontaneously breaks time diffeomorphism invariance, giving rise to a Goldstone boson $\pi$ that transforms as $\pi(x) \rightarrow \pi(x) - \xi(x)$ under time-diffeomorphisms $t \rightarrow t + \xi(x)$.  The degrees of freedom in the effective theory are $\pi$, the metric fluctuations, and the fluctuations of the $N$ isocurvaton fields $\sigma_I$.  To construct the effective theory, one writes all possible terms invariant under the full set of space-time diffeomorphisms involving these fields (see \cite{Noumi:2012vr} for a complete derivation in the context of single-isocurvaton QSFI).  

We will be interested in computing the momentum space correlation functions of $\pi$ at the time when the modes exit the horizon.  This implies we can work in the ``decoupling limit" of the effective theory \cite{Cheung:2007st} and set the metric perturbations to zero.  
Correlation functions of $\zeta$ in the gauge where $\pi$ is zero can be related to correlation functions of $\pi$ in the decoupling limit by \cite{Cheung:2007st}
\begin{align}\label{curv to pi}
  \zeta = -(H/\dot{\phi}) \pi.
\end{align}
The leading quadratic Lagrangian simplifies to any term quadratic in $\pi$ and $\sigma_I$ that is consistent with a shift symmetry in $\pi$ and a background de-Sitter space-time:
\begin{align}
\label{L2}
    {\cal L}_2 = \frac{1}{2(H \tau)^2}&\left(\left(\partial_\tau \pi\right)^2-\tilde{c}_\pi^2\left(\partial_i \pi\right)^2+\tilde{Z}_{IJ}\partial_\tau \sigma_I \partial_\tau \sigma_J-\tilde{c}^{2}_{\sigma IJ}\partial_i \sigma_I \partial_i \sigma_J+2\tilde{\beta}_{I}\partial_\tau \pi \partial_\tau \sigma_I + \tilde{\delta}_{I} \partial_i \pi \partial_i \sigma_I\right.\cr 
    &\left.-\left(H\tau\right)^{-2}\tilde{m}^{2}_{IJ}\sigma_I \sigma_J -2\left(H \tau\right)^{-1}\tilde{\rho}_{IJ} \sigma_I\partial_\tau \sigma_J  - 2\left(H \tau\right)^{-1}\tilde{\mu}_I \sigma_I\partial_\tau \pi \right).
\end{align}
Note, repeated indices are summed over and we have included the $\sqrt{-g}$ factor from the action in the Lagrangian.  We have dropped terms quadratic in the fields that have more than two derivatives because they are suppressed by powers of the cutoff $\Lambda$ of the effective theory.  

Several terms in (\ref{L2}) can be eliminated by field redefinitions.  For example, the $\tilde \delta_{I}\partial_i\pi\partial_i \sigma_I$ interaction can be absorbed into other couplings by performing the time diffeomorphism that induces the shift $\pi \rightarrow \pi + \tilde \delta_I \sigma_I/(2 c^2_\pi)$.  Moreover, we can rotate and re-scale $\sigma_I$ to diagonalize $\tilde c^{2}_{\sigma IJ}$ and set $\tilde Z_{I J } = 1$.  Equation (\ref{L2}) then simplifies to 
\begin{align}\label{L2 simple}
    {\cal L}_2 = \frac{1}{2(H\tau)^2}\big( & \left(\partial_\tau \pi\right)^2-c_\pi^2\left(\partial_i \pi\right)^2+\left(\partial_\tau \sigma_I\right)^2-c_{\sigma I}^2\left(\partial_i \sigma_I\right)^2+2\beta_I \partial_\tau \pi \partial_\tau \sigma_I \cr
    &  -\left(H \tau \right)^{-2} m_{IJ}^2\sigma_I \sigma_J-2\left(H\tau\right)^{-1}\rho_{IJ} \sigma_I \partial_\tau \sigma_J -2\left(H \tau\right)^{-1}\mu_I \sigma_I\partial_\tau \pi\big).
\end{align}
The matrix $m^2_{IJ}$ is symmetric while $\rho_{IJ}$ is anti-symmetric\footnote{Any symmetric part of $\rho_{IJ}$ can be absorbed into $ m^2_{IJ}$ through an integration by parts.}.  The interactions $\mu_I\dot{\pi}\sigma_I$ and $\rho_{IJ}\sigma_I \dot \sigma_J$ could result from a UV theory containing terms such as $(\sigma_I/\Lambda) g^{\mu\nu}\partial_{\mu}\phi\partial_{\nu}\phi$ and $(\sigma_I/\Lambda)g^{\mu\nu}\partial_{\mu} \sigma_J \partial_{\nu}\phi$.  The kinetic mixings in (\ref{L2 simple}) arise because the inflaton vev spontaneously breaks Lorentz invariance.  If $\dot{\phi}_{0} = 0$, Lorentz invariance is unbroken and the kinetic mixings must vanish.  By dimensional analysis, this means $\mu_{I}$ and $\rho_{IJ}$ are proportional to $\dot{\phi}_{0}/\Lambda$.  Using $(H^2/\dot{\phi}_{0})^2 \sim 2\pi^2\Delta_{\zeta}^2$, where 
\begin{equation}
    \Delta_\zeta^2=\frac{k^3}{2\pi^2}P_\zeta(k)\simeq 2.11\times 10^{-9}
\end{equation}
is the dimensionless power spectrum \cite{Aghanim:2018eyx}, the cutoff can be expressed in terms of $|\mu_I|$ as
\begin{align}
    \frac{\Lambda}{H} \sim \frac{H}{|\mu_I|}\times 10^4.
\end{align}
For $\mu_I \sim O(H)$, this implies that $\Lambda\gg H$ and higher derivative terms are suppressed.

We can recover previously studied single-isocurvaton QSFI models by taking limits of (\ref{L2 simple}).  If we take $N=1$, $\beta_1 = 0$ and $c^2_\pi=c^2_{\sigma1}=1$, we recover the quadratic part of the QSFI Lagrangian originally considered by Chen and Wang \cite{Chen:2009zp}.  The resulting Lagrangian only has two parameters, $\mu_1$ and $m_{11}$.  Single-isocurvaton QSFI with generic speeds of sound and nonzero $\beta_1$ was studied in \cite{Noumi:2012vr}.  The presence of a nontrivial $\rho_{IJ}$ matrix is the main new aspect of theories with $N>1$ isocurvatons.

One way to treat the kinetic mixings parameterized by $\mu_I$, $\beta_I$ and $\rho_{IJ}$ is to write the Fourier transforms of the $\pi$ and $\sigma_I$ fields in terms of a common set of raising and lowering operators
\begin{align}\label{mode expansion}
\hat{\pi}(\tau,{\bf x}) &= \int \frac{d^{3}k}{(2\pi)^3}\frac{H}{k^{3/2}}\sum_{i=1}^{N+1}\left(\hat{a}_{\bf{k}}^{(i)}\pi^{(i)}(\eta)e^{-i\bf{k}\cdot\bf{x}} + c.c\right)\cr
\hat{\sigma}_I(\tau,\bf{x}) &= \int \frac{d^{3}k}{(2\pi)^3}\frac{H}{k^{3/2}}\sum_{i=1}^{N+1}\left(\hat{a}_{\bf{k}}^{(i)}\sigma_{I}^{(i)}(\eta)e^{-i\bf{k}\cdot\bf{x}} + c.c\right)
\end{align}
where $\eta = k\tau$.  The mode functions $\pi^{(i)}(\eta)$ and $\sigma^{(i)}(\eta)$ obey the Euler-Lagrange equations obtained from (\ref{L2 simple}),
\begin{align}
\label{mode EOM}
&{\sigma^{(i)}_I}''-\frac{2}{\eta}{\sigma^{(i)}_I}'+c^2_{\sigma I}{\sigma^{(i)}_I}+\frac{m^2_{IJ}}{\eta^2}{\sigma^{(i)}_J} +\frac{\mu_I}{\eta}{\pi^{(i)}}'+\beta_I\left({\pi^{(i)}}''-\frac{2}{\eta}{\pi^{(i)}}'\right)+\frac{\rho_{IJ}}{\eta}\left(2{\sigma^{(i)}_J}'-\frac{3}{\eta}{\sigma^{(i)}_J}\right)= 0\cr
&{\pi^{(i)}}''-\frac{2}{\eta}{\pi^{(i)}}'+c^2_\pi {\pi^{(i)}}-\frac{\mu_I}{\eta}\left({\sigma^{(i)}_I}' - \frac{3}{\eta}{\sigma^{(i)}_I}\right)+\beta_I\left({\sigma^{(i)}_I}''-\frac{2}{\eta}{\sigma^{(i)}_I}'\right) = 0
\end{align}
where primes denote derivatives with respect to $\eta$.  The mode functions asymptotically obey the Bunch-Davies vacuum condition.  Since equations (\ref{mode EOM}) are coupled, they are difficult to solve analytically for general parameters\footnote{Analytic progress can be made in the context of single-isocurvaton QSFI in the regimes where $\mu_1$, $m_{11} \ll H$ \cite{An:2017rwo}, or $\mu_1$, $m_{11} \gg H$ \cite{Baumann:2011su,An:2017hlx,Gwyn:2012mw,Assassi:2013gxa}.}.  Instead, we use the numerical solutions of (\ref{mode EOM}) to perform most of the calculations in our analysis.  However, one can derive the small $\eta$ behavior of the mode functions analytically, which turns out to fix the wavevector dependence of the squeezed and collapsed limits of the bispectrum and trispectrum.

In the limit\footnote{We have chosen the convention in which inflation occurs between $-\infty < \tau \le 0$} $-\eta << 1$, we can neglect the terms in (\ref{mode EOM}) proportional to the speeds of sound.  The leading late time behavior for the mode functions can be written as
\begin{align}
\label{IR mode}
    \pi^{(i)}(\eta) &= a^{(i)}_s (-\eta)^s\ \ \ \ \sigma^{(i)}_I(\eta) = b^{(i)}_{I,s} (-\eta)^s.
\end{align}
For example, specializing to $N = 2$ and inserting (\ref{IR mode}) into (\ref{mode EOM}) yields the following equation for $s$,
\begin{align}
\begin{vmatrix}
    (s-3)s       &\  (s-3)(s\beta_1-\mu_1) &\  (s-3)(s\beta_2-\mu_2)\\
    s((s-3)\beta_1 + \mu_1)       &\  m_{11}^2+(s-3)s &\  m_{12}^2+(3-2 s)\rho\\
    s((s-3)\beta_2 + \mu_2)       &\  m_{12}^2+(2s - 3)\rho &\  m_{22}^2 + (s-3)s
\end{vmatrix} = 0,
\end{align}
which can be solved for six different roots:
\begin{align}\label{s roots}
    s = 0,s_-,s_-^{*},3-s_-,3-s_-^{*},3.
\end{align}
In general, $s_-$ is a complex number satisfying\footnote{For general model parameters $\text{Re}[s_-]$ can be less than 0, however the modes would then be tachyonic and grow rather than decay as $\eta \rightarrow 0$.} $0<\text{Re}[s_-]\le 3/2$.  The $s=0$ solution arises from the shift symmetry in $\pi$ and can only exist in the $\pi$ mode functions.  The leading $\eta$ behavior of the mode functions is 
\begin{align}
\label{late mode}
    \pi^{(i)}(\eta) &= a^{(i)}_0+a^{(i)}_{s_-}(-\eta)^{s_-} + a^{(i)}_{s_-^*}(-\eta)^{s_-^*}+\dots\cr \sigma^{(i)}_I(\eta) &= b^{(i)}_{I,s_-}(-\eta)^{s_-} + b^{(i)}_{I,s_-^*}(-\eta)^{s_-^*}+\dots.
\end{align}
Equations (\ref{late mode}) imply that as $\eta \rightarrow 0$, $\pi^{(i)}$ approaches a constant while $\sigma^{(i)}_I$ decays to $0$. 

The late time behavior of $\sigma_I$ in multi-isocurvaton QSFI can be very different from its late time behavior in single-isocurvaton QSFI \cite{Chen:2009zp}.  If we write $s_- \equiv \alpha + i\gamma$, equation (\ref{late mode}) becomes
\begin{align}
    \sigma^{(i)}_I(\eta) = b^{(i)}_{I,s_{-}}(-\eta)^\alpha e^{i\gamma {\log}(-\eta)} + b^{(i)}_{I,s_{-}^{*}}(-\eta)^\alpha e^{-i\gamma {\log}(-\eta)}.
\end{align}
Observe that the modes oscillate logarithmically in $\eta$ with frequency $\gamma$.  Moreover, $\alpha$ dictates how quickly the modes decay at late times.  In the original Chen and Wang theory \cite{Chen:2009zp} $\gamma$ can only be nonzero when $\alpha = 3/2$ (see e.g. \cite{An:2017hlx}).  This means that while the isocurvaton's mode functions exhibit oscillatory behavior at late times, they decay quickly as $\eta \rightarrow 0$.  On the other hand, in a multi-isocurvaton QSFI theory with $\rho_{IJ} \neq 0$, one can obtain $\gamma \neq 0$ with $\alpha < 3/2$, which means that $\sigma^{(i)}_I$ can oscillate while decaying slowly.

To illustrate this, we specialize to the case of two isocurvatons and focus on two sets of parameters which will serve as our benchmark models.  The first set is
\begin{align}
\label{first set}
  &\ \ \mu_1 = m_{12} = m_{21} = \beta_I =0,\ c_{\pi}= c_{\sigma_1} = c_{\sigma_2} = 1,\cr
    &\mu_2 = 0.6H,\ m_{11}^2 = m_{22}^2 =-\rho_{12}^2 = -(5H)^2
\end{align}
which yields $s_{-} \simeq 0.06 - 5.00 i$, while the second is
\begin{align}
\label{second set}
&\ \ \ \mu_1 = m_{12} = m_{21} = \beta_I =0,\ c_{\pi}= c_{\sigma_1} = c_{\sigma_2} = 1,\cr
    &\mu_2 = 0.4H,\ m_{11} = m_{22} = 0.3 H,\ \rho_{12} = H
\end{align}
which yields $s_{-} \simeq 0.46 - 1.00i$.  Notice, the masses squared in (\ref{second set}) are negative, which is usually a signal of tachyonic modes whose mode functions diverge as $\eta \rightarrow 0$.  However, due to the kinetic mixing, the mass squared parameters that appear in the Lagrangian do not equal the physical masses squared.  Indeed, $\alpha > 0$ for this set of parameters, which implies $\sigma_I \rightarrow 0$ as $\eta\rightarrow 0$.

For $\alpha<3/2$ and $\gamma\neq 0$, $\alpha$ and $\gamma$ are typically the same order of magnitude.  Some tuning is required in order to produce $\alpha\ll 1$ with $\gamma \sim O(1)$.  This means, rapid oscillations that decay slowly cannot be produced without some degree of tuning between model parameters\footnote{For example, in eq. (\ref{second set}), we tuned $m_{ii}^2=-\rho^2$.}.

\section{Primordial Non-Gaussianity}
\label{nongaussianity section}
In the previous section, we showed that theories with multiple-isocurvaton  have a  kinetic mixing term parameterized by the matrix $\rho_{IJ}$ that cannot exist in single-isocurvaton models.  If this term is present, the mode functions of $\pi$ and $\sigma_I$ can exhibit oscillatory behavior that decays slowly at late times.  We now study the effects of this behavior on the non-Gaussianities of the scalar curvature perturbations of the metric, $\zeta$.  

We are interested in computing the ``in-in" correlation functions of $\zeta$ at $\tau = 0$, which are related to those of $\pi$ through (\ref{curv to pi}).  
The in-in correlator of an operator $O$ can be expressed in the so-called ``commutator form" (see for example  \cite{Weinberg:2005vy}) as
\begin{equation}\label{in in}
    \langle {\cal O}(0)\rangle = \sum_{N=0}^\infty i^N\int_{-\infty}^{0} d {\tau_N} \int_{-\infty}^{\tau_N} d\tau_{N-1}\dots\int_{-\infty}^{\tau_2} d\tau_1\langle [H_\text{int}(\tau_1),[H_\text{int}(\tau_2),\dots [H_\text{int}(\tau_N),{\cal O}(0)]\dots]]\rangle_I
\end{equation}
where the fields on the right hand side of (\ref{in in}) evolve according to (\ref{mode expansion}) and (\ref{mode EOM}).  In general, the interaction Hamiltonian consists of an isocurvaton potential $V(\sigma_I)$ as well as interactions involving combinations of $\pi$ and $\sigma_I$.  For simplicity, we assume the potential consists of a cubic interaction involving only the $\sigma_1$ field and that the interaction Hamiltonian is dominated by this interaction\footnote{The potential $V(\sigma_I)$ is not related to the other operators in the effective theory by any symmetry and can, in principle, be the largest term in the interaction Hamiltonian.}:
\begin{equation}\label{Hint}
    H_\text{int}(\tau) = \frac{1}{(H \tau)^4}\int d^3x\frac{V'''}{3!}\sigma_1(x)^3 + \dots.
\end{equation}
In the next two subsections, we compute the bispectrum and trispectrum of $\zeta$ due to this interaction.
\subsection{Bispectrum}

\begin{figure}
\centering
\includegraphics[width=1.5in]{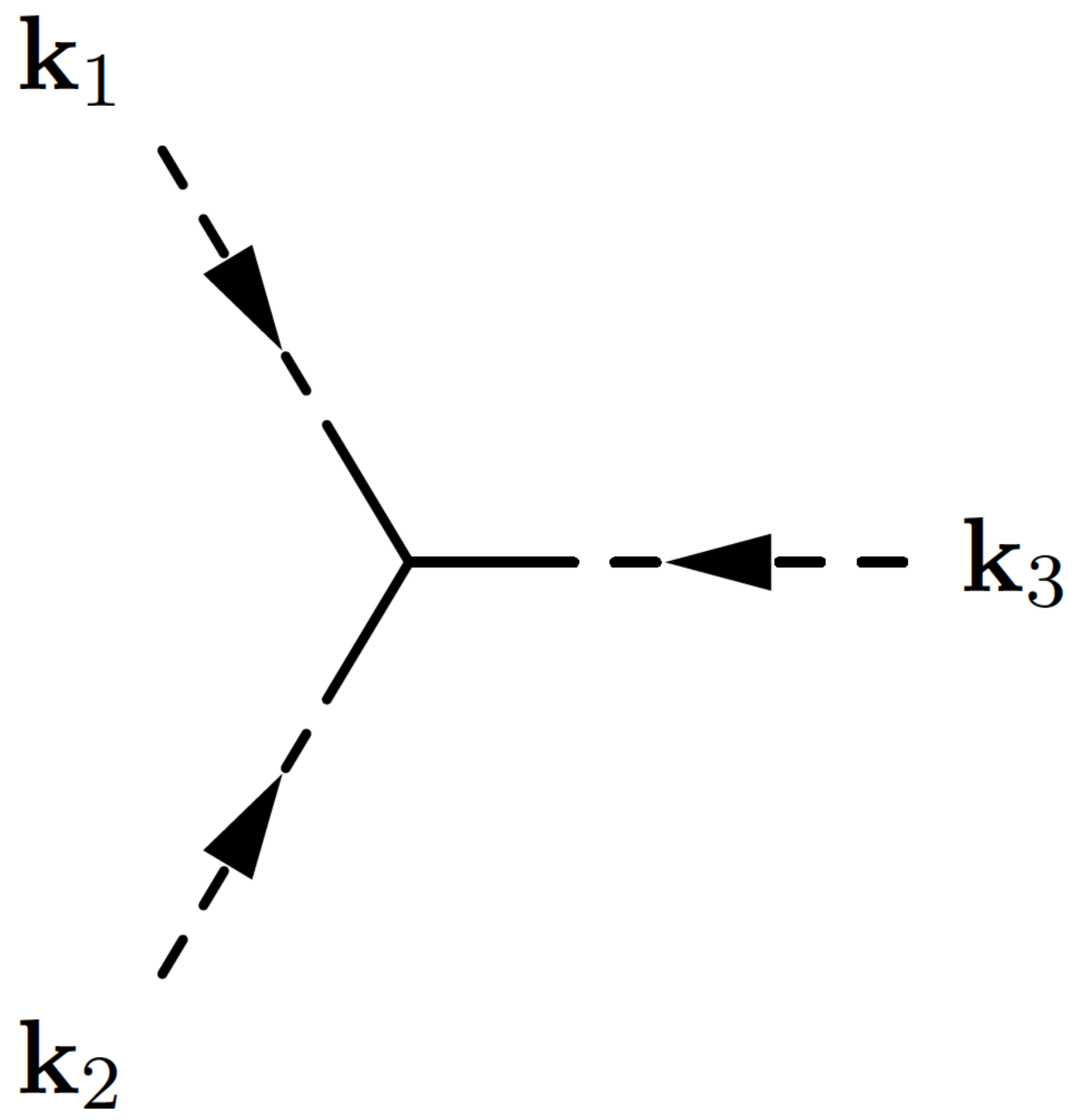}
\caption{A diagramatic representation of the leading contribution to the bispectrum.  Dashed lines represent $\pi$, while solid lines represent $\sigma_1$.}
\label{fig:threepoint}
\end{figure}

We first consider the three-point function of $\zeta$.  Let $\zeta_{\bf k}$ denote the Fourier transform of $\zeta({\bf x},0)$.  The bispectrum $B_\zeta$ is defined as
\begin{equation}
    \langle \zeta_{{\bf k}_1} \zeta_{{\bf k}_2} \zeta_{{\bf k}_3}\rangle = B_\zeta({\bf k}_1,{\bf k}_2,{\bf k}_3)(2\pi)^3\delta^3({\bf k}_1+{\bf k}_2+{\bf k}_3).
\end{equation}
Using equations (\ref{curv to pi}), (\ref{mode expansion}), (\ref{in in}) and (\ref{Hint}), it is straightforward to show that to leading order in $V'''$
\begin{align}\label{bispectrum}
    B_\zeta(k_1,k_2,k_3) &= -2\left(\frac{H^2}{\dot\phi_0}\right)^3\frac{V'''}{H}\frac{1}{k_1^3k_2^3k_3^3}\cr
    & \times\int_{-\infty}^0\frac{d\tau}{\tau^4}{\rm Im}\left(\pi^{(i)}(0)\sigma^{(i)*}_1(k_1\tau)\pi^{(j)}(0)\sigma^{(j)*}_1(k_2\tau)\pi^{(k)}(0)\sigma^{(k)*}_1(k_3\tau)\right)
\end{align}
where repeated mode labels are summed.  Note, the sum $\pi^{(i)}(0)\sigma_1^{(i)*}(k_i\tau)$ is nonzero because of the kinetic mixings.

The $\tau$ integral in (\ref{bispectrum}) is potentially IR divergent due to the factor of $1/\tau^4$ in the integrand.  Even though we do not have explicit expressions for the mode functions, it can be shown using the canonical commutation relations (see Appendix \ref{commutator constraints}) that the integral is indeed finite in the IR.  One can then evaluate the bispectrum using the numerical solutions of (\ref{mode EOM}). 

Consider the squeezed limit of (\ref{bispectrum}), which occurs when $k_l\equiv k_1\sim k_2$ and $k_s \equiv k_3 \ll k_l$, \textit{i.e.} there are two long sides and one short side of the triangle traced out by the ${\bf k}_i$.  We can factor out the momentum dependence from the time integral by changing integration variables to $\eta = k_l\tau$ and expanding to leading order in $k_s/k_l$ using (\ref{late mode}).  We find
\begin{align}\label{squeezed}
    B_\zeta^{\text{sq}}(k_l,k_s)&= -4\left(\frac{H^2}{\dot\phi_0}\right)^3\left(\frac{V'''}{H}\right)\frac{1}{k_l^6}\left(\frac{k_s}{k_l}\right)^{-3+\alpha}\cr
    \Big(&\cos\left(\gamma\log k_s/k_l\right)\text{Re}\left[a^{(i)}_0b^{(i)*}_{1,s_-}y^*(\alpha,\gamma)\right]\cr
    +&\sin\left(\gamma\log k_s/k_l\right)\text{Im}\left[a^{(i)}_0b^{(i)*}_{1,s_-}y^*(\alpha,\gamma)\right] \Big)
\end{align}
where
\begin{align}\label{y integrals}
    y(\alpha,\gamma)&=\int_{-\infty}^0\frac{d\eta}{(-\eta)^{4-\alpha-i \gamma}}\text{Im}\left[\left(\pi^{(i)}(0)\sigma^{(i)*}_1(\eta)\right)^2\right].
\end{align}
The squeezed bispectrum oscillates logarithmically in $k_s/k_l$ with angular frequency $\gamma$ and the amplitude grows as $(k_s/k_l)^{-3+\alpha}$.  As mentioned earlier, in single-isocurvaton QSFI, $\gamma$ can only be nonzero when $\alpha = 3/2$, which means the amplitude of an oscillating squeezed bispectrum can only grow as $(k_s/k_l)^{-3/2}$ in these models.  This is not the case for the multi-isocurvaton models given by (\ref{first set}) or (\ref{second set}), whose oscillating bispectrum grow approximately as $(k_s/k_l)^{-2.94}$ and $(k_s/k_l)^{-2.54}$ respectively.  

\begin{figure}[t]
    \centering
    \includegraphics[width=6in]{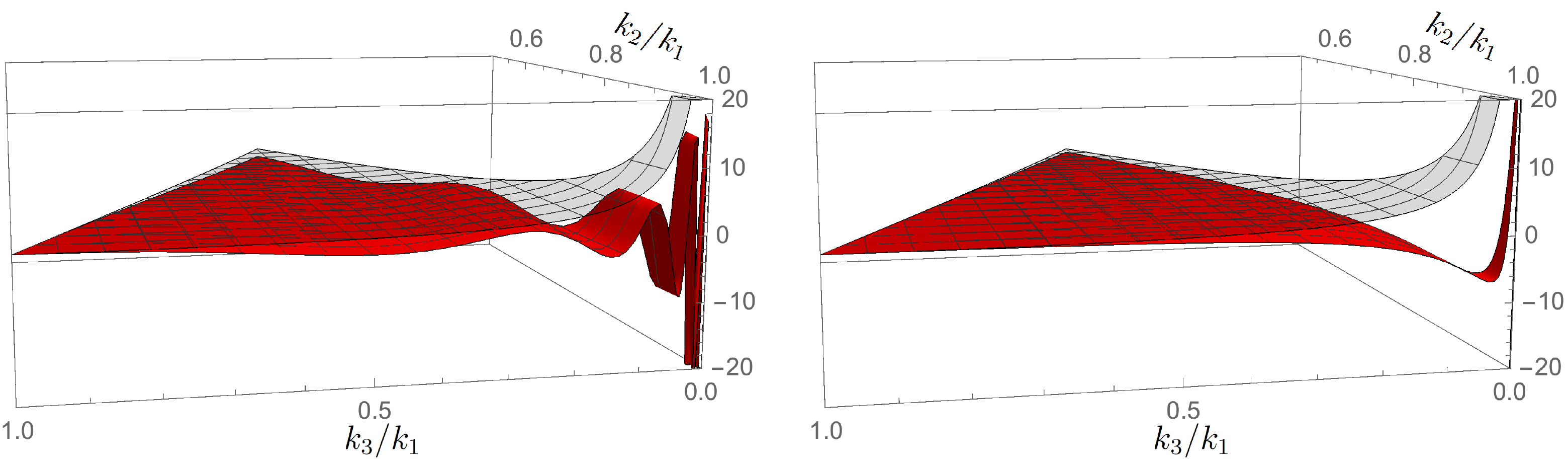}
    \caption{Plots of the shape functions $S(k_1,k_2,k_3)$ in the range $k_1>k_2>k_3$ for multi-isocurvaton oscillatory shape (red) and local shape (transparent gray).  In the left and right panels we plot parameters (\ref{first set}) and (\ref{second set}) respectively.}
    \label{fig:bisp figure 1}
\end{figure}

 To illustrate the momentum dependence of the full bispectrum, it is useful to define the shape function
\begin{align}
    S(k_1,k_2,k_3)=\kappa(k_1k_2k_3)^2B_\zeta(k_1,k_2,k_3).
\end{align}
  The normalization factor $\kappa$ is chosen so that $S(k,k,k) = 1$.

In Fig.~\ref{fig:bisp figure 1}, we plot the shape function of multi-isocurvaton QSFI for the model parameters (\ref{first set}) and (\ref{second set}).  For comparison, we also include the shape function of local non-Gaussianity
\begin{align}
    S^{\rm loc}(k_1,k_2,k_3)=\frac{1}{3}(k_1k_2k_3)^2\left[\frac{1}{(k_1k_2)^3}+\frac{1}{(k_2k_3)^3}+\frac{1}{(k_3k_1)^3}\right]
\end{align}
which is close to the shape function of the single-isocurvaton QSFI originally considered by Chen and Wang in the limit $\mu_{1},m_{11} \ll H$ \cite{Chen:2009zp,An:2017rwo}. 

Fig.~\ref{fig:bisp figure 1} is consistent with the analytic results for the squeezed bispectrum of multi-isocurvaton QSFI.  In the limit $k_3/k_1 \rightarrow 0$, the shape function oscillates logarithmically in $k_3/k_1$ and its amplitude has power law growth.  On the other hand, in single-isocurvaton QSFI, the shape function for an oscillating bispectrum decays to 0 as $k_3/k_1\rightarrow 0$.  Note, to get very rapid oscillations and nearly local power law growth for the multi-isocurvaton shape, one needs to tune parameters as in (\ref{first set}).  However, even the shape for the untuned parameters (\ref{second set}) displays a visible turn due to nonzero $\gamma$.

\begin{figure}
    \centering
    \includegraphics[width=6in]{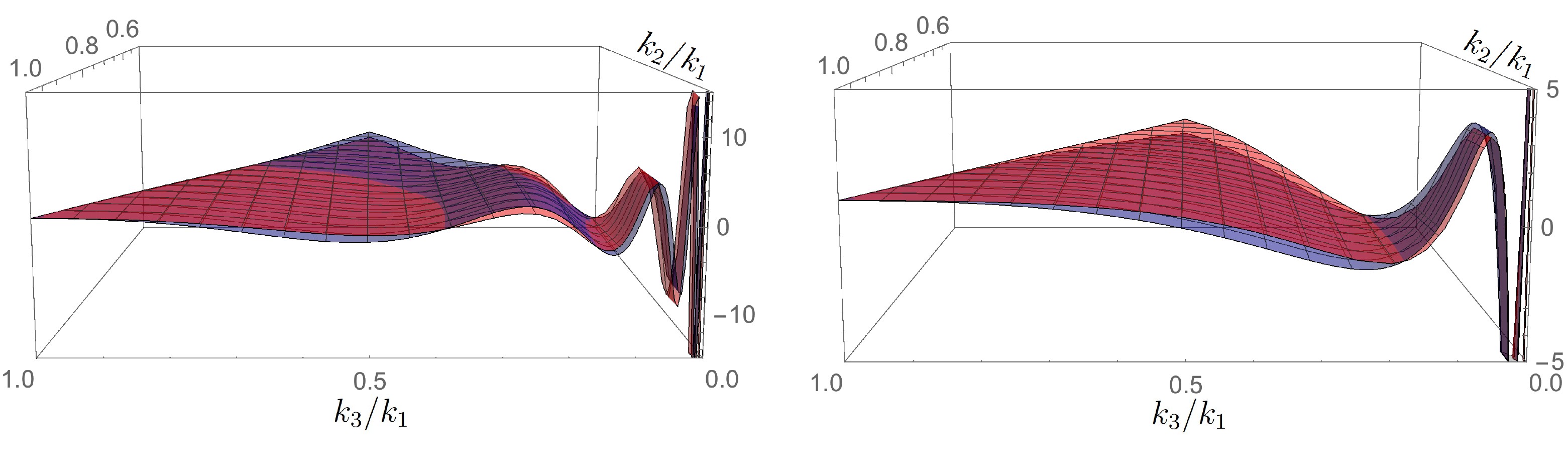}
    \includegraphics[width=6in]{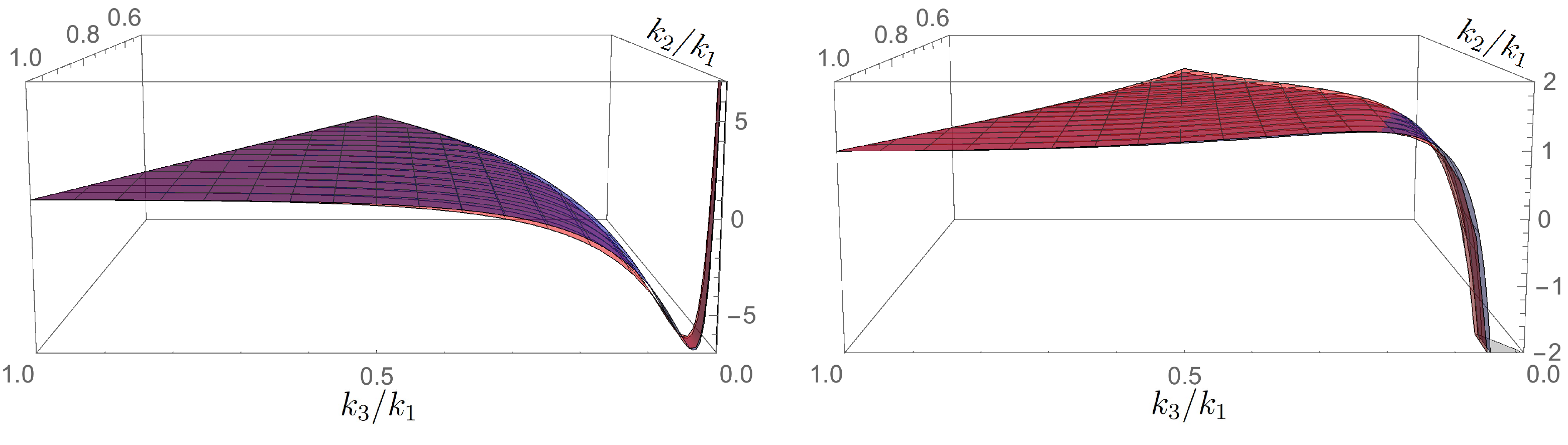}
    \caption{We plot shape functions $S(k_1,k_2,k_3)$ of the multi-isocurvaton oscillatory shape in the range $k_1>k_2>k_3$.  We plot the shape computed numerically in red and plot the approximate shape defined in eq. (\ref{the shape}) in blue.  The top-left plot has parameters $(\alpha,\gamma,\phi)=(0.06,-5.00,-1.55)$, top-right $(0.23,-2.50,-1.58)$, bottom-left $(0.46,-1.00,3.08)$, and bottom-right $(0.14,-0.50,-2.78)$. }
    \label{fig:shapeaccuracyfig}
\end{figure}

For single-isocurvaton QSFI with $\gamma=0$, the shape function has been approximated as \cite{Chen:2009zp}:
\begin{equation}
    S_\alpha^{\rm QSFI}(k_1,k_2,k_3)=3^{3\alpha-2}\frac{\left(k_1^2+k_2^2+k_3^2\right)\left(k_1+k_2+k_3\right)^{1-3\alpha}}{\left(k_1k_2k_3\right)^{1-\alpha}}
\end{equation}
A good phenomenological fit to the multi-isocurvaton QSFI shape is
\begin{equation}\label{the shape}
    S_{\alpha,\gamma,\phi}^{\rm OQSFI}(k_1,k_2,k_3)=C\left(\cos\left(\gamma \log\left(\frac{k_2 k_3}{k_1\left(k_1+k_2+k_3\right)}\right)+\phi \right)+\text{2 perms.}\right)S_\alpha^{\rm QSFI}(k_1,k_2,k_3)
\end{equation}
where the normalization $C=\left(3\cos\left(\gamma \log\left(1/3\right)+\phi\right)\right)^{-1}$ enforces $S^{\rm OQSFI}=1$ in the equilateral limit.  Note that the shape function is parameterized by three numbers $\alpha$, $\gamma$, and a phase $\phi$.  In Fig. \ref{fig:shapeaccuracyfig}, we plot the shape functions evaluated numerically against the shape functions computed with (\ref{the shape}).  

We can also define the parameter $f_{\rm NL}^{\rm OQSFI}$, corresponding to the magnitude of this shape.  In keeping with convention\footnote{Conventionally, $f_{\rm NL}$ is defined via $B_\Phi(k_1,k_2,k_3)=6A^2f_{\rm NL}\frac{1}{k_1^2k_2^2k_3^2}S(k_1,k_2,k_3)$ where $A$ is given by the power spectrum's normalization $P_\Phi(k)=A/k^3$ (see, e.g. \cite{Ade:2015xua}).}, we define:
\begin{equation}\label{fnl def}
    B_\zeta(k_1,k_2,k_3)=\frac{18}{5}(k^3P_\zeta(k))^2 f_{\rm NL}^{\rm OQSFI}\frac{1}{k_1^2k_2^2k_3^2}S_{\alpha,\gamma,\phi}^{\rm OQSFI}(k_1,k_2,k_3).
\end{equation}
By matching the squeezed limit of (\ref{fnl def}) onto (\ref{squeezed}), we can obtain $V'''/H$ as a function of $f_{\rm NL}^{\rm OQSFI}$:
\begin{equation}
    \frac{V'''}{H}=-\frac{9}{10}(|a^{(i)}_0|^2)^{3/2}\sqrt{2\pi^2 \Delta_\zeta^2}f_{\rm NL}^{\rm OQSFI}\frac{C}{\sqrt{\text{Im}\left[a^{(i)}_0b^{(i)*}_{1,s_-}y^*(\alpha,\gamma)\right]^2+\frac{1}{9}\text{Re}\left[a^{(i)}_0b^{(i)*}_{1,s_-}y^*(\alpha,\gamma)\right]^2 }}.
\end{equation}

Even for $\gamma\sim O(0.1)$, the oscillating QSFI bispectrum shape still displays qualitatively distinct features. Specifically, the shape can get large and negative in the squeezed limit.  This is different from many other bispectrum shapes that grow in the squeezed limit because they typically remain positive rather than become negative (see \cite{Ade:2015ava}).

\subsection{Collapsed Trispectrum}
\begin{figure}
\centering
\includegraphics[width=2.8in]{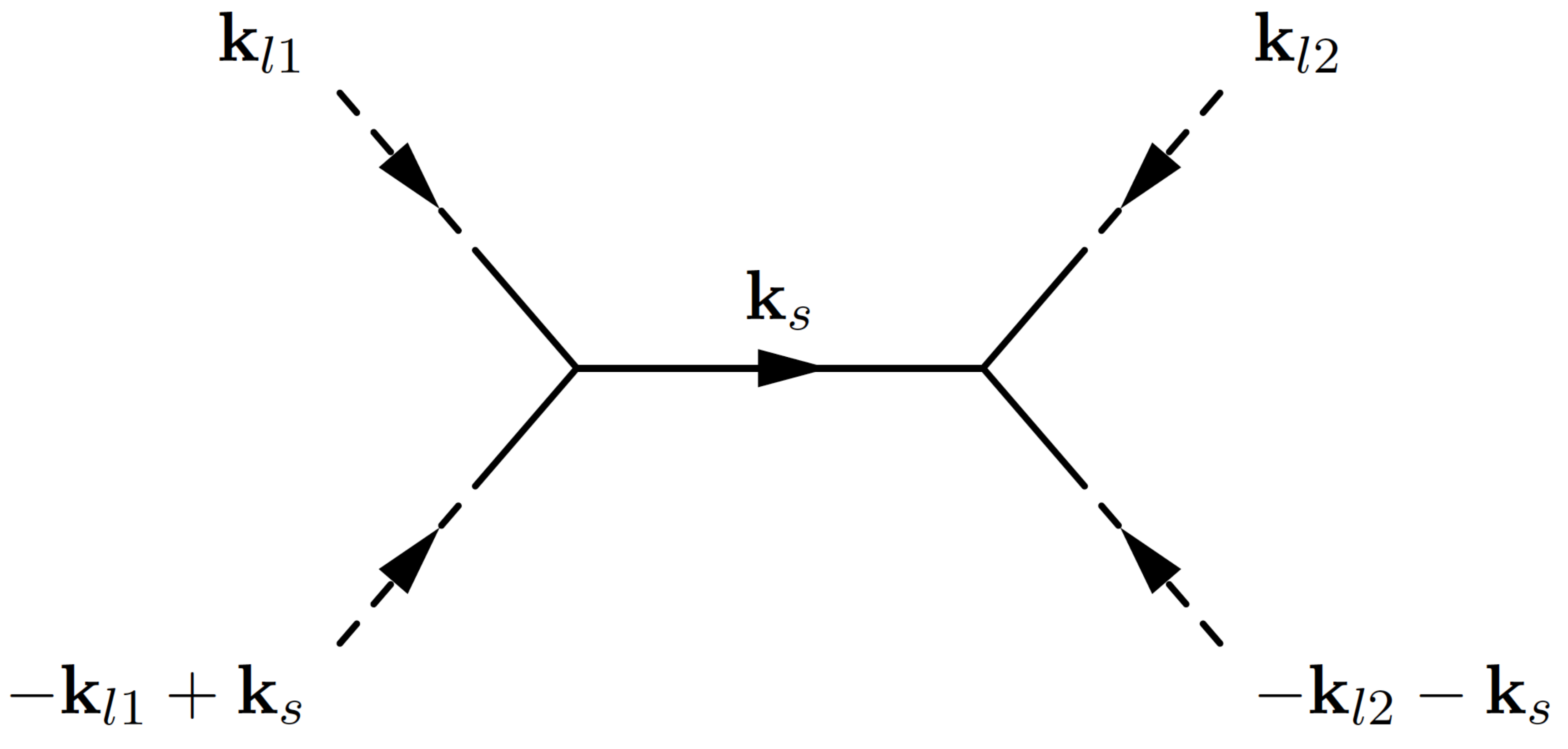}
\label{fig:fourpoint}
\caption{A diagramatic representation of the collapsed trispectrum.  Dashed lines represent $\pi$, while solid lines represent $\sigma_1$.}
\end{figure}
We now consider the four-point function of $\zeta$, which is defined as
\begin{equation}
    \langle \zeta_{{\bf k}_1} \zeta_{{\bf k}_2} \zeta_{{\bf k}_3}\zeta_{{\bf k}_4}\rangle = T_\zeta({\bf k}_1,{\bf k}_2,{\bf k}_3,{\bf k}_4)(2\pi)^3\delta^3({\bf k}_1+{\bf k}_2+{\bf k}_3+{\bf k}_4).
\end{equation}
We focus on the collapsed limit, which occurs when $k_{l1}\equiv k_1\sim k_2$, $k_{l2}\equiv k_3\sim k_4$, and $k_s\equiv |{\bf k}_1+{\bf k}_2|=|{\bf k}_3+{\bf k}_4|\ll k_{li}$.  The integrand will be symmetric in $k_{l1}$ and $k_{l2}$, which means we can untangle the limits of the time integrals.  The collapsed four-point then simplifies to
\begin{align}\label{collapsed trispectrum}
    {T_\zeta}^\text{coll}(k_{l1},k_{l2},k_s)=4&\left(\frac{H^2}{\dot\phi_0}\right)^4\left(\frac{V'''}{H}\right)^2\frac{1}{k_{l1}^{9/2}k_{l2}^{9/2}}\left(\frac{k_s^2}{k_{l1}k_{l2}}\right)^{-3/2+\alpha}\cr
    \bigg[&\cos\left(\gamma\log \frac{k_{l2}}{k_{l1}}\right)\left(|b^{(i)}_{1,s_-}|^2+|b^{(i)}_{1,s_-^*}|^2\right)\left|y(\alpha,\gamma)\right|^2\cr
    +2&\cos\left(\gamma\log\frac{k_s^2}{k_{l1}k_{l2}}\right)\text{Re}\left[b^{(i)}_{1,s_-} b^{(i)*}_{1,s_-^*}y(\alpha,\gamma)^2\right]\cr
    -2&\sin\left(\gamma\log\frac{k_s^2}{k_{l1}k_{l2}}\right)\text{Im}\left[b^{(i)}_{1,s_-} b^{(i)*}_{1,s_-^*}y(\alpha,\gamma)^2\right]\bigg].
\end{align}
Note, not every term oscillates logarithmically in $k_s/k_l$.  Instead, there are two terms which oscillate with angular frequency $2\gamma$ as well as one that does not oscillate at all in $k_s/k_l$.  

It turns out that a loop contribution to the trispectrum can produce terms that oscillate with frequency $4\gamma$.  We explore this in Appendix \ref{loops}.

\section{Large-Scale Structure}
\label{lls section}
In this section, we determine the effects of PNG on large-scale structure, specifically the halo-halo power spectrum $P_{hh}$ and the matter-halo power spectrum $P_{hm}$.  It is well known that in certain inflationary theories, such as single-isocurvaton QSFI,  the contributions of PNG to $P_{hh}$ and $P_{hm}$ can become much larger than the Gaussian contributions at wavevectors of order $(10^2 \text{ Mpc}/h)^{-1}$ \cite{An:2017rwo, McAneny:2017bbv,An:2018tcq}.  We now consider this in multi-isocurvaton QSFI.  

The primordial curvature perturbations are related to the linearly evolved smoothed matter density perturbations today, $\delta_{R}(\bf{k})$, by \cite{Dodelson book}
\begin{equation}\label{matter curvature}
    \delta_R({\bf k})=\frac{2k^2}{5\Omega_m H_0^2}T(k)W_R(k)\zeta_{\bf k}
\end{equation}
where $\Omega_m$ is the ratio of the matter density and the critical density today, $H_0$ is the Hubble constant today, $T(k)$ is the BBKS transfer function \cite{Bardeen:1985tr} and $W_R(k)$ is a window function smoothing over radius $R$.  We use the top-hat window function 
\begin{align}
    W_{R}(k) = \frac{3 \left({\rm sin}kR - kR {\rm cos}kR\right)}{\left(kR\right)^3}.
\end{align}.

Since we are interested in scales of order $10^2$ Mpc/$h$, $\delta_h$ can be related to $\delta_R$ through a bias expansion.  For simplicity, we use a local-in-matter-density bias expansion (for a more comprehensive treatment, see \cite{Desjacques:2016bnm})
\begin{equation}\label{bias_expansion}
    \delta_h({\bf x})=b_1\delta_R({\bf x})+b_2\left(\delta_R({\bf x})^2-\langle\delta_R({\bf x})^2\rangle\right)+\dots
\end{equation}
The bias coefficients can be approximated using the threshold model introduced in \cite{Kaiser:1984sw}.  We assume that halos form instantaneously at some redshift $z_{\rm coll}$, and that halos only form at points where the overdensity exceeds some critical threshold $\delta_c(z_{\rm{coll}})$.  We also neglect the evolution of halos after collapse.  In this model, the bias coefficients $b_1$ and $b_2$ are then
\begin{align}
    b_1 = 2\frac{e^{-\delta_c^2/(2\sigma_R^2)}}{\sqrt{2\pi}\sigma_R{\rm \ erfc}(\delta_c/(\sqrt{2}\sigma_R))}  && b_2 =  \frac{e^{-\delta_c^2/(2\sigma_R^2)}\delta_c}{\sqrt{2\pi}\sigma_R^3{\rm \ erfc}(\delta_c/(\sqrt{2}\sigma_R))}
\end{align}
where  $\sigma_R^2 = \langle  \delta_R({\bf x})^2\rangle $.   In deriving our numerical results, we use $\delta_c = 4.215$, which corresponds to $\delta_c(z_{\rm coll})=1.686$ at $z_{\rm coll} = 1.5$ \cite{Press:1973iz}, and $R = 3 {\rm \ Mpc}/h$.
 While a more sophisticated treatment of halo dynamics will change our precise numerical results, we do not expect them to impact our conclusions qualitatively.

We will be interested in computing the halo-halo power spectrum $P_{hh}$, 
\begin{equation}\label{halo halo}
    \langle \delta_h({\bf k}_1)\delta_h({\bf k}_2)\rangle = P_{hh}(k_1)(2\pi)^3\delta^3({\bf k}_1+{\bf k}_2)
\end{equation}
as well as the halo-matter power spectrum $P_{hm}$,
\begin{equation}\label{halo matter}
    \langle \delta_h({\bf k}_1)\delta_{R}({\bf k}_2)\rangle = P_{hm}(k_1)(2\pi)^3\delta^3({\bf k}_1+{\bf k}_2).
\end{equation}
The Gaussian contributions to $P_{hh}$ and $P_{hm}$ are found to be
\begin{align}
    &P_{hh}(k_s)\Big|_G=b_1^2\left(\frac{H^2}{\dot\phi_0}\right)^2\left(\frac{2}{5\Omega_m H_0^2 R^2}\right)^2k_sR^4|a^{(i)}_0|^2\cr 
    &P_{hm}(k_s)\Big|_G=b_1\left(\frac{H^2}{\dot\phi_0}\right)^2\left(\frac{2}{5\Omega_m H_0^2 R^2}\right)^2k_sR^4|a^{(i)}_0|^2.
\end{align}
At wavevectors of order $(10^2 \text{ Mpc}/h)^{-1} \ll R^{-1}$, the most significant non-Gaussian contributions to $P_{hh}$ are due to the squeezed and collapsed limits of the bispectrum (\ref{squeezed}) and trispectrum (\ref{collapsed trispectrum}).  Then, plugging (\ref{matter curvature}) and  (\ref{bias_expansion}) into (\ref{halo halo}) and using (\ref{squeezed}) and (\ref{collapsed trispectrum}) gives
\begin{align}\label{halo halo ratio}
    \frac{P_{hh}(k_s)}{P_{hh}(k_s)\big|_G}=1&-8\left(\frac{b_2}{b_1}\right)\left(\frac{H^2}{\dot\phi_0}\right)\left(\frac{2}{5\Omega_m H_0^2 R^2}\right)\left(\frac{V'''}{H}\right)\frac{1}{(k_sR)^{2-\alpha}}\frac{1}{|a^{(i)}_0|^2}\cr
    & \ \ \ \ \ \bigg(\cos\left(\gamma\log k_sR\right)\text{Re}\left[J^*(\alpha,\gamma)y^*(\alpha,\gamma)a^{(i)}_0b^{(i)*}_{1,s_-}\right]\cr
    & \ \ \ \ \ \ +\sin\left(\gamma\log k_sR\right)\text{Im}\left[J^*(\alpha,\gamma)y^*(\alpha,\gamma)a^{(i)}_0b^{(i)*}_{1,s_-}\right]\bigg)\cr
    &+8\left(\frac{b_2}{b_1}\right)^2\left(\frac{H^2}{\dot\phi_0}\right)^2\left(\frac{2}{5\Omega_m H_0^2 R^2}\right)^2\left(\frac{V'''}{H}\right)^2\frac{1}{(k_sR)^{4-2\alpha}}\frac{1}{|a^{(i)}_0|^2}\cr
    & \ \ \ \ \ \bigg(\left|J(\alpha,\gamma)\right|^2\left|y(\alpha,\gamma)\right|^2|b^{(i)}_{1,s_-}|^2\cr
    & \ \ \ \ \ \ +\cos\left(2\gamma\log k_sR\right)\text{Re}\left[J(\alpha,\gamma)^2y(\alpha,\gamma)^2b^{(i)}_{1,s_-}b^{(i)*}_{1,s_-^*}\right]\cr
    & \ \ \ \ \ \ -\sin\left(2\gamma\log k_sR\right)\text{Im}\left[J(\alpha,\gamma)^2y(\alpha,\gamma)^2b^{(i)}_{1,s_-}b^{(i)*}_{1,s_-^*}\right]\bigg).
\end{align}
where, to compactify notation, we have defined
\begin{align}
    J(\alpha,\gamma)&=\frac{1}{2\pi^2}\int_0^\infty dx \  x^{3-\alpha-i \gamma}T(x/R)^2W(x/R)^2
\end{align}
The most significant non-Gaussian contribution to $P_{hm}$ comes from the squeezed bispectrum:
\begin{align}
     \frac{P_{hm}(k_s)}{P_{hm}(k_s)\big|_G}=1&-4\left(\frac{b_2}{b_1}\right)\left(\frac{H^2}{\dot\phi_0}\right)\left(\frac{2}{5\Omega_m H_0^2 R^2}\right)\left(\frac{V'''}{H}\right)\frac{1}{(k_sR)^{2-\alpha}}\frac{1}{|a^{(i)}_0|^2}\cr
    & \ \ \ \ \ \bigg(\cos\left(\gamma\log k_sR\right)\text{Re}\left[J^*(\alpha,\gamma)y^*(\alpha,\gamma)a^{(i)}_0b^{(i)*}_{1,s_-}\right]\cr
    & \ \ \ \ \ \ +\sin\left(\gamma\log k_sR\right)\text{Im}\left[J^*(\alpha,\gamma)y^*(\alpha,\gamma)a^{(i)}_0b^{(i)*}_{1,s_-}\right]\bigg).
\end{align}
\begin{figure}[t]
    \centering
    \includegraphics[width=6.5in]{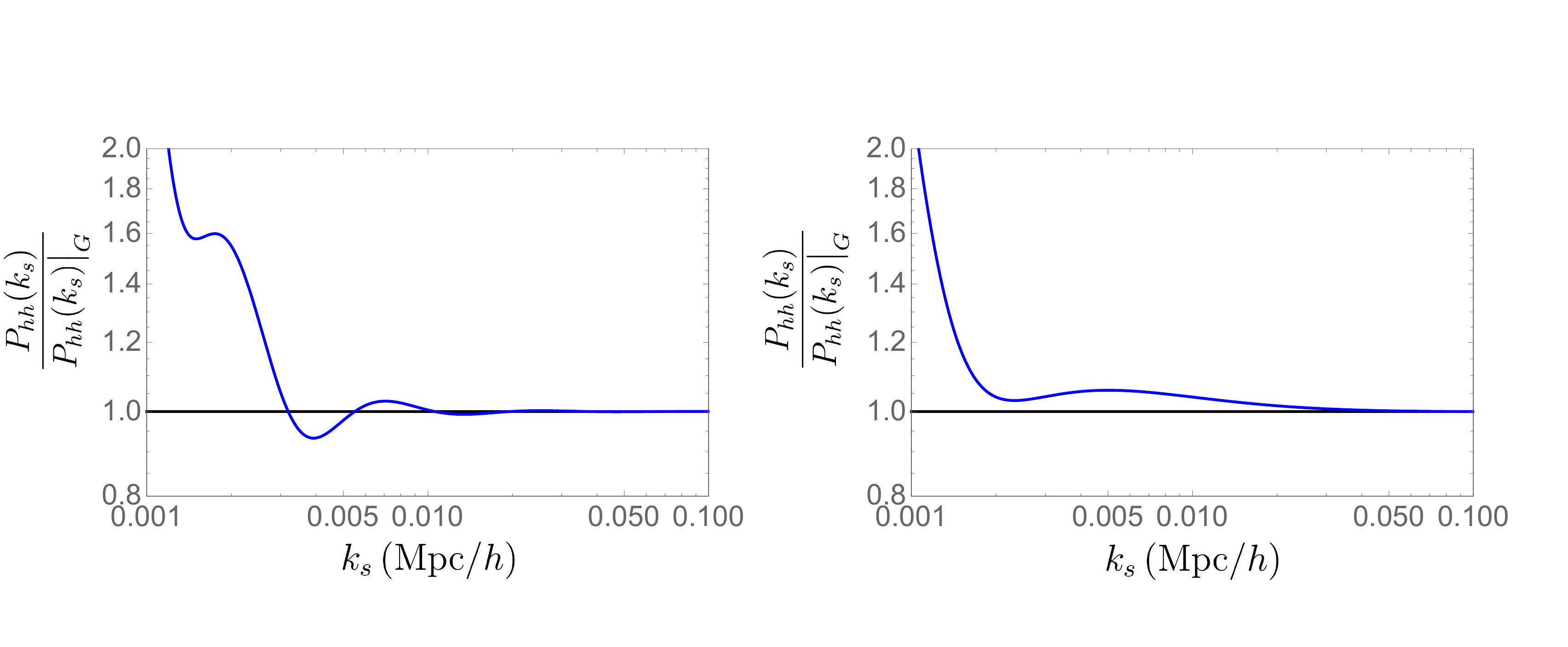}
    \caption{We plot the halo-halo power spectrum scaled by the Gaussian halo-halo power spectrum (i.e. the power spectrum for $V'''=0$).  In the left panel, we plot the parameters from eq. (\ref{first set}) with $f_{\rm NL}^{OQSFI}=10$, and in the right panel, we plot the parameters from eq. (\ref{second set}) with $f_{\rm NL}^{OQSFI}=-10$.}
\label{fig:halo halo plots}
\end{figure}

In Figure \ref{fig:halo halo plots} we plot (\ref{halo halo ratio}) for the model parameters (\ref{first set}) and (\ref{second set}) and $|f_{\rm NL}^{\rm OQSFI}|=10$.  Note that at around $(10^{2} {\rm \ Mpc}/h)^{-1}$, $P_{hh}$ begins to deviate from $P_{hh}|_G$.  The oscillations evident in Fig.~\ref{fig:halo halo plots} are a consequence of the oscillatory squeezed bispectrum and collapsed trispectrum.

However, for $|f_{\rm NL}^{\rm OQSFI}|\sim 10$, the amplitude of the oscillations is quite small. Moreover, the non-Gaussian contributions to $P_{hh}$ only begin to dominate over the Gaussian contribution at a scale of order $(10^3 {\rm \ Mpc}/h)^{-1}$, which is unlikely to be detected experimentally in the near future. 

This scale is smaller than the scale at which the non-Gaussian contributions to $P_{hh}$ begin to dominate in single-isocurvaton QSFI theories with $f_{NL} \sim 10$ and $\alpha$ small \cite{An:2017rwo}.  The reason is the integrals involving the transfer functions, $J(\alpha,\gamma)$, are oscillatory in multi-isocurvaton QSFI when $\gamma \neq 0$, which washes them out.  This makes the coefficients of the non-Gaussian contributions smaller in multi-isocurvaton QSFI than in single-isocurvaton QSFI.

\section{Conclusion}
In this paper, we studied quasi-single field inflation with multiple isocurvatons.  
Multi-isocurvaton QSFI includes the interaction $\rho_{IJ}\dot{\sigma}_I \sigma_J$ which can give rise to novel inflationary dynamics.  In particular, the mode functions of $\pi$ and $\sigma_I$ can exhibit late time log oscillations that decay slowly as $\eta \rightarrow 0$.  Due to these late time oscillations, the primordial non-Gaussianities of $\zeta$ exhibit log-oscillatory behavior in ratios of wavevector magnitudes.

For example, the bispectrum is proportional to $(k_s/k_l)^{-3+\alpha}\cos(\gamma \log k_s/k_l)$ in the squeezed limit, which means for small $\alpha$ it experiences nearly cubic growth while oscillating.  This behavior cannot be achieved in single-isocurvaton QSFI.  
Furthermore, the collapsed trispectrum goes as $(k_s/k_l)^{-3+2\alpha}(a + b\cos(2\gamma \log k_s/k_l))$, \textit{i.e.} there is a term that does not oscillate in $k_s/k_l$ as well as one that does with frequency $2\gamma$.

In models where $\alpha \lesssim 0.5$, the contributions of the squeezed bispectrum and collapsed trispectrum to the halo-halo power spectrum $P_{hh}(k_s)$ and the halo-matter power spectrum $P_{hm}(k_s)$ can dominate over the Gaussian contributions at $k_s \sim (10^3 {\rm \ Mpc}/h)^{-1}$. When $\gamma\neq0$, $P_{hh}$ and $P_{hm}$ oscillate logarithmically in $k_s R$.

\section*{ACKNOWLEDGEMENTS}

We would like to thank Mark Wise for useful discussions. This work was supported by the DOE Grant DE-SC0011632. We are
also grateful for the support provided by the Walter Burke Institute for Theoretical Physics.

\appendix

\section{Commutator Constraints}
\label{commutator constraints}
The $\eta$ integrals of (\ref{y integrals}) are potentially IR divergent because of the factors of $1/\eta^4$ in the integrands.  It can be shown that all potentially IR divergent terms are zero.  Eq.~(\ref{mode EOM}) implies that the $\sigma$ mode functions can be written in series form as
\begin{align}\label{series form}
\sigma_I^{(i)}(\eta) &= b^{(i)}_{I,s_-}(-\eta)^{s_-} + b^{(i)}_{I,s_-^*}(-\eta)^{s_-^*} + b^{(i)}_{I,2}(-\eta)^2 + b^{(i)}_{I,2+s_-}(-\eta)^{2+s_-} + b^{(i)}_{I,2+s_-^*}(-\eta)^{2+s_-^*}\cr
    &\ \ \ \ \ \ \ \ \ \ \ \ \ \ \ + b^{(i)}_{I,3-s_-}(-\eta)^{3-s_-} + b^{(i)}_{I,3-s_-^*}(-\eta)^{3-s_-^*} + b^{(i)}_{I,3} (-\eta)^3 + \dots    
\end{align}
The equal time commutation relation $\left[\pi,\sigma_I\right] = 0$ holds order by order in powers of $\eta$ and implies the following relations among the power series coefficients
\begin{align}\label{series coeff relations}
&\pi^{(i)}(0)b^{(i)*}_{I,s_-} = \left(\pi^{(i)}(0)b^{{(i)}*}_{I,s_-^*} \right)^*,\ \  \pi^{(i)}(0)b_{I,s_-}^{{(i)}*} =  \left(\pi^{(i)}(0)b_{I,s^{*}_{-}}^{(i)*} \right)^*\nonumber\\
& \ \ \pi^{(i)}(0)b_{I,3-s_-}^{(i)*} =  \left(\pi^{(i)}(0)b_{I,3-s^{*}_{-}}^{(i)*} \right)^*, \ \ {\rm Im}\left(\pi^{(i)}(0)b^{(i) *}_{I,2}\right) = 0
\end{align}
where the sum over mode label $i$ is implicit.  Furthermore, the mode equations (\ref{mode EOM}) imply
\begin{align}\label{coeff relation}
b^{(i)}_{I,2+s_{-}} = c b^{(i)}_{I ,s_{-}}\ \ \ \ b^{(i)}_{I,2+s^{*}_{-}} = c^
*  b^{(i)}_{I,s^{*}_{-}}.
\end{align}
Combining the first relation in (\ref{series coeff relations}) with (\ref{coeff relation}) yields
\begin{align}
    \pi^{(i)}(0)b^{(i)*}_{I,2+s_-} = \left(\pi^{(i)}(0)b^{(i)*}_{I,2+s_-^*} \right)^*
\end{align}
The leading infrared behavior of (\ref{y integrals}) is then
\begin{align}\label{y IR forms}
y(\alpha,\gamma) &= 2\ {\rm Im}\left[\pi^{(i)}(0)b^{(i)*}_{1,3} \right]\int_{-\infty}^0 \frac{d\eta}{(-\eta)^{1-\alpha-i\gamma}} {\rm Re}\left[\pi^{(j)}(0)\sigma^{(j)*}_1(\eta)\right].
\end{align}
It is straightforward to fit the $\sigma^{(i)}_1$ mode functions to (\ref{series form}) and determine the numerical coefficient in (\ref{y IR forms}).  The method to evaluate $y(\alpha,\gamma)$ is to choose an $-\eta_{IR} < 1$ and numerically integrate from $-\infty < \eta < \eta_{IR}$ using (\ref{y integrals}) and then integrate from $\eta_{IR} \leq \eta < 0$ using (\ref{y IR forms}).

\section{Loop Contribution to the Collapsed Trispectrum}\label{loops}

Recall that the tree-level collapsed trispectrum, which has one internal line, has terms that oscillate with frequency $2\gamma$.  We now show that a loop contribution to the collapsed trispectrum, which has two internal lines, can contain terms that oscillate with frequency $4\gamma$.  The loop contribution to the collapsed trispectrum then induces terms in the halo-halo power spectrum that oscillate as ${\rm cos} (4 \gamma{\rm log}k_s R)$.\footnote{Contributions to the halo-halo power spectrum due to primordial non-Gaussianities sourced by quantum loops in the context of single-isocurvaton QSFI were considered in \cite{McAneny:2017bbv,An:2018tcq}.}

\begin{figure}
\centering
\includegraphics[width=2.8in]{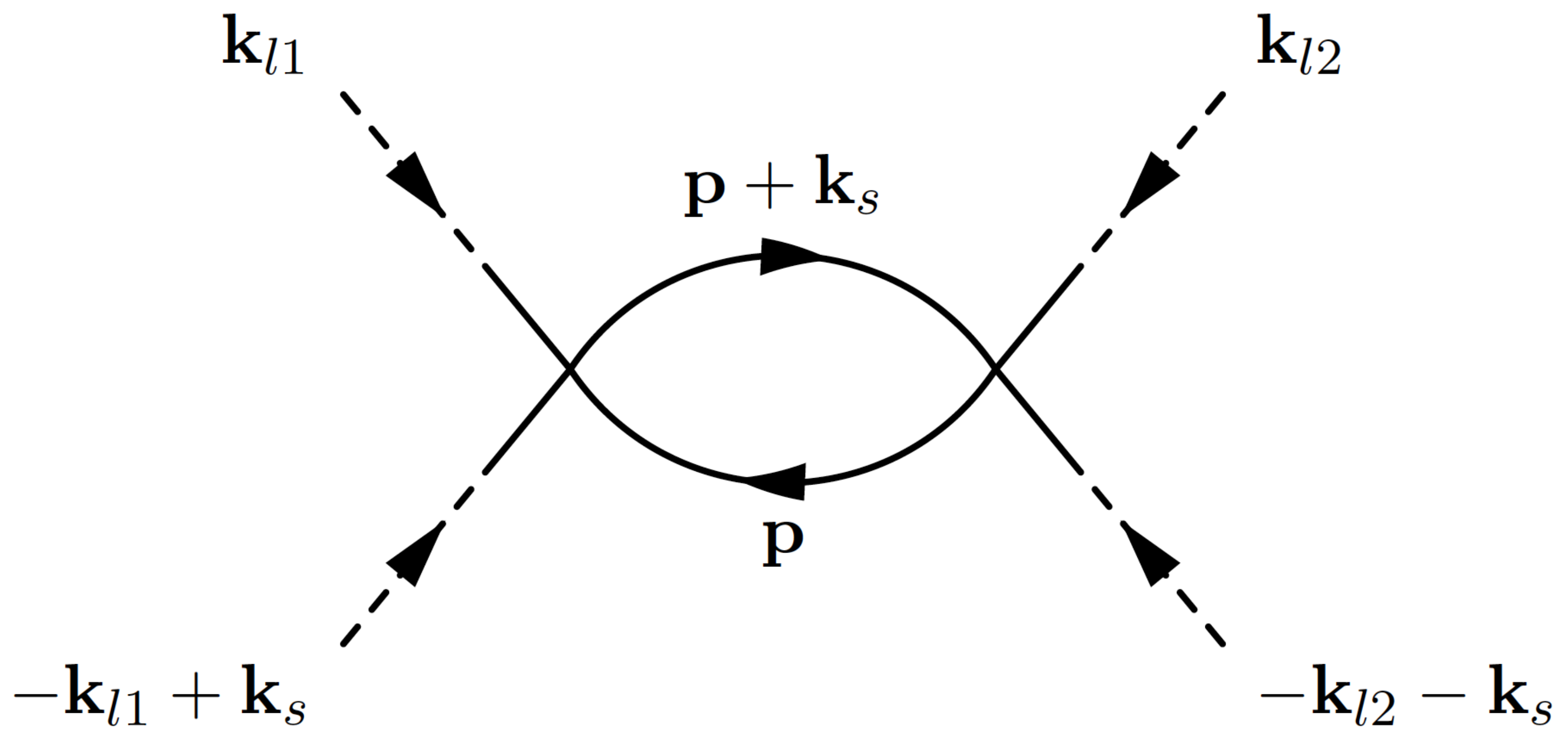}
\label{fig:loop}
\caption{A diagramatic representation of the loop contribution to the collapsed trispectrum.  Dashed lines represent $\pi$, while solid lines represent $\sigma_1$.}
\end{figure}
Consider a theory in which the interaction Hamiltonian is composed of a single $\sigma_1^4$ interaction:
\begin{equation}\label{loop interaction hamiltonian}
    H_\text{int}(\tau) = \frac{1}{(H \tau)^4}\int d^3 x\frac{V''''}{4!}\sigma_1(x)^4.
\end{equation}
Inserting two factors of (\ref{loop interaction hamiltonian}) into (\ref{in in}) yields the 1-loop contribution to the trispectrum:
\begin{align}\label{full loop cont}
    T_\zeta^\text{coll}(k_{l1},&k_{l2},k_s) = 4 \left(\frac{H^2}{\dot \phi_0}\right)^4{V''''}^2\frac{1}{k_{l1}^3k_{l2}^3}\int_{-\infty}^0\frac{d\eta}{\eta^4}\int_{-\infty}^{\frac{k_{l2}}{k_{l1}}\eta}\frac{d\eta'}{\eta'^4}\int \frac{d^3p}{(2\pi)^3}\frac{1}{p^3|{\bf p}+{\bf k_s}|^3}\nonumber\\
    &\times\text{Im}\left[\left(\pi^{(i)}(0)\sigma_1^{(i)*}(\eta)\right)^2\right] \text{Im}\left[\left(\pi^{(j)}(0)\sigma_1^{(j)*}(\eta')\right)^2\left(\sigma_1^{(l)}\left(\frac{p}{k_{l1}}\eta\right)\sigma_1^{(l)*}\left(\frac{p}{k_{l2}}\eta'\right)\right)\right.\nonumber\\
    &\left.\ \ \ \ \ \ \ \ \ \ \ \ \ \ \ \ \ \ \ \ \ \ \ \ \ \ \ \ \ \ \ \ \ \ \ \ \ \ \ \times\left(\sigma_1^{(m)}\left(\frac{|{\bf p}+{\bf k_s}|}{k_{l1}}\eta\right)\sigma_1^{(m)*}\left(\frac{|{\bf p}+{\bf k_s}|}{k_{l2}}\eta'\right)\right)\right]
\end{align}
As described in \cite{Arkani-Hamed:2015bza,McAneny:2017bbv,An:2018tcq}, for $k_{s}\ll k_l$, the loop diagram can give a large contribution to the collapsed trispectrum.  In this limit, the loop integral is dominated by the region in which $p \sim k_s$, which means the mode functions on the bottom line of (\ref{full loop cont}) can be expanded in $p/k_{l i}$.  Defining the integral
\begin{equation}
    Z(\epsilon_1,\epsilon_2)\equiv \int\frac{d^3u}{(2\pi)^3}u^{-3+2\epsilon_1}|{\bf u}+{\bf \hat x}|^{-3+2\epsilon_2},
\end{equation}
where ${\bf \hat x}$ is an arbitrary unit vector, the loop contribution can be written:
\begin{align}\label{loop trispectrum}
    {T_\zeta}^\text{coll}(&k_{l1},k_{l2},k_s)= \ 2\left(\frac{H^2}{\dot\phi_0}\right)^4{V''''}^2\frac{1}{k_{l1}^{9/2}k_{l2}^{9/2}}\left(\frac{k_s^2}{k_{l1}k_{l2}}\right)^{-3/2+2\alpha} \ \ \ \ \ \ \ \ \ \ \ \ \ \ \ \ \ \ \ \ \ \ \ \ \ \ \ \ \ \ \ \ \ \ \ \ \ \ \ \cr
    & \bigg[\text{Re}[c_1]+\text{Re}[c_2]\cos\left(2\gamma\log \frac{k_{l2}}{k_{l1}}\right)+\text{Re}[c_3] \cos\left(2\gamma\log \frac{k_s}{k_{l1}}\right)+\text{Re}[c_3] \cos\left(2\gamma\log \frac{k_s}{k_{l2}}\right)\cr
    &-\text{Im}[c_3]\sin\left(2\gamma\log \frac{k_s}{k_{l1}}\right)-\text{Im}[c_3] \sin\left(2\gamma\log \frac{k_s}{k_{l2}}\right)+\text{Re}[c_4] \cos\left(2\gamma\log\frac{k_s^2}{k_{l1}k_{l2}}\right)\cr&-\text{Im}[c_4]\sin\left(2\gamma\log\frac{k_s^2}{k_{l1}k_{l2}}\right)\bigg]
\end{align}
where
\begin{align}
    c_1 &= 2y(2\alpha,0)^2\left(|b^{(i)}_{1,s_-}|^2|b^{(j)}_{1,s_-^*}|^2Z(\alpha,\alpha)+|b^{(i)}_{1,s_-}b^{(i)*}_{1,s_-^*}|^2Z(\alpha+i\gamma,\alpha-i \gamma)\right)\cr
    c_2 &=\left|y(2\alpha,2\gamma)\right|^2\left(\left(|b^{(i)}_{1,s_-}|^2\right)^2+\left(|b^{(i)}_{1,s_-^*}|^2\right)^2\right)Z(\alpha,\alpha)\cr
    c_3 & = 4y(2\alpha,0)y(2\alpha,2\gamma)b^{(i)}_{1,s_-}b^{(i)*}_{1,s_-^*}\left(|b^{(j)}_{1,s_-}|^2+|b^{(j)}_{1,s_-^*}|^2\right)Z(\alpha,\alpha+i \gamma)\cr
    c_4 & = 2\left|y(2\alpha,2\gamma)\right|^2\left(b^{(i)}_{1,s_-}b^{(i)*}_{1,s_-^*}\right)^2Z(\alpha+i\gamma,\alpha+i\gamma).
\end{align}
Note that the loop contribution to the trispectrum has terms that do not oscillate with $\log k_s/k_{li}$, terms that oscillate with frequency $2\gamma$, and terms that oscillate with frequency $4\gamma$.

The loop contribution to $P_{hh}$ due to the quartic $\sigma_1^4$ interaction can also be computed:
\begin{align}\label{loop lss}
    \frac{P_{hh}(k_s)\big|_\text{loop}}{P_{hh}(k_s)\big|_G} \ = \ & 2\left(\frac{b_2}{b_1}\right)^2\left(\frac{H^2}{\dot \phi_0}\right)^2\left(\frac{2}{5\Omega_mH_0^2 R^2}\right)^2{V''''}^2\frac{1}{(k_s R)^{4-2\alpha}}\frac{1}{|a^{(i)}_0|^2} \ \ \ \ \ \ \ \ \ \ \ \ \ \ \ \ \ \ \ \ \ \ \cr
    &\Big(\text{Re}\left[c_1 J(2\alpha,0)^2+c_2\left|J(2\alpha,2\gamma)\right|^2\right]\cr
    &+2\cos\left(2\gamma \log k_s R\right)\text{Re}\left[ c_3J(2\alpha,0)J(2\alpha,2\gamma)\right]\cr
    &-2\sin\left(2\gamma \log k_s R\right)\text{Im}\left[ c_3J(2\alpha,0)J(2\alpha,2\gamma)\right]\cr
    &+\cos\left(4\gamma \log k_s R\right)\text{Re}\left[ c_4J(2\alpha,2\gamma)^2\right]\cr
    &-\sin\left(4\gamma \log k_s R\right)\text{Im}\left[ c_4J(2\alpha,2\gamma)^2\right]\Big)
\end{align}

While it appears that the terms oscillating with frequency $4\gamma$ in (\ref{loop lss}) could induce unique observable features in $P_{hh}(k_s)$, it turns out their coefficients are suppressed relative to those of the non-oscillating terms since $|J(2\alpha,2\gamma)|\ll J(2\alpha, 0)$.  This can be understood by noting that while the magnitudes of the integrands of $J(2\alpha,2\gamma)$ and $J(2\alpha, 0)$ are the same, the integrand of $J(2\alpha,2\gamma)$ is oscillatory and washes $J(2\alpha,2\gamma)$ out.

\end{document}